\documentclass[twocolumn,prl,showpacs,nofootinbib,superscriptaddress]{revtex4}
\usepackage{latexsym}
\usepackage{dcolumn}
\usepackage{epsfig}
\RequirePackage{graphicx}
\def\lpa{\lambda_{p{-}\rm air}}
\def\spa{\sigma_{p{-}\rm air}}
\def\spai{\sigma_{p{-}\rm air}^{\rm prod}}
\def\spae{\sigma_{p{-}\rm air}^{\rm el}}
\def\spaqe{\sigma_{p{-}\rm air}^{q{-}\rm el}}

\newcommand{\ba}{\begin{eqnarray}}

\newcommand{\ea}{\end{eqnarray}}
\newcommand{\be}{\begin{equation}}
\newcommand{\ee}{\end{equation}}

\newcommand{\eq}[1]{Eq.\,(\ref{#1})}

\def\bea{\begin{eqnarray}} 
\def\eea{\end{eqnarray}} 
\begin{document}

\preprint{ANL-HEP-PR-06-79}

\title{Ultra-high Energy Predictions of proton-air Cross Sections from Accelerator Data}

\author{M.~M.~Block}
\affiliation{Department of Physics and Astronomy, Northwestern University, 
Evanston, IL 60208}

\begin{abstract}
We predict $\spai$, the  proton--air inelastic production 
 cross section, at $pp$ center-of-mass energies $2\le\sqrt s \le 100000$ TeV, using high energy predictions from a saturated Froissart bound parameterization of accelerator data on forward $\bar pp$ and $pp$ scattering amplitudes, together with Glauber theory.   The parameterization of the $\bar pp$ and $pp$ cross sections incorporates analyticity constraints and unitarity, allowing accurate extrapolations to ultra-high energies. Our predictions are in excellent agreement with cosmic ray extensive air shower measurements, both in magnitude and in energy dependence. 
\end{abstract}

\pacs{13.60.Hb, 12.38.-t, 12.38.Qk}

\maketitle

{\em Introduction.} There are now available published p-air inelastic production cross sections\cite{fly,akeno,yakutsk,eastop} ($\spai$) that span the enormous $pp$ cms (center-of-mass system) energy range $2\le \sqrt s\le 100000$ TeV, reaching energies  well above the Large Hadron Collider (LHC).  Moreover, there are also now available very accurate predictions at cosmic ray energies for the total $pp$ cross section, $\sigma_{pp}$,  from fits\cite{blockhalzenpp} to accelerator data that used adaptive data sifting algorithms\cite{sieve} and analyticity constraints{\cite{blockanalyticity}.  However, extracting proton--proton cross sections from published cosmic ray observations of extensive air showers, and vice versa, is far from
 straightforward\cite{engel}. By a variety of experimental techniques,
 cosmic ray experiments map the atmospheric depth at which extensive air 
 showers develop and measure the distribution of $X_{\rm max}$, the shower maximum, which is sensitive to the inelastic p-air cross section $\spai$. From the measured $X_{\rm max}$ distribution, the experimenters deduce $\spai$. In this note we will compare published values of $\spai$ with predictions made from  $\sigma_{pp}$, using a Glauber model to obtain $\spai$ from $\sigma_{pp}$.
  
{\em $\spai$ from the $X_{\rm max}$ distribution: Method I.} The measured shower attenuation length ($\Lambda_m$) is not only
 sensitive to the interaction length of the protons in the atmosphere
 ($\lpa$), with
\begin{equation}
\Lambda_m = k \lpa = k { 14.4 m_p \over \spai}=k\frac{24,100}{\spai} \,,  \label{eq:Lambda_m}
\end{equation}
(with $\Lambda_m$ and $\lpa$ in g\,cm$^{-2}$, the proton mass $m$ in g, and the inelastic production cross section $\spai$ in mb),  but also depends on the rate at which the energy of the primary proton
 is dissipated into electromagnetic shower energy observed in the
 experiment. The latter effect is parameterized in Eq.\,(\ref{eq:Lambda_m})
 by the parameter $k$. The value of $k$ depends critically on the inclusive
 particle production cross section and its energy dependence in nucleon and meson interactions
 on the light nuclear target of the atmosphere (see Ref. \cite{engel}). We emphasize that the goal of the cosmic ray experiments is $\spai$ (or correspondingly, $\lpa$), whereas 
 in Method I, the {\em measured} quantity is $\Lambda_m$. Thus,  
a significant drawback of Method I is that one needs a model of
 proton-air interactions to complete the loop between the measured
 attenuation length $\Lambda_m$ and the cross section $\spai$,
 {\em i.e.,} one needs the value of $k$ in Eq. (\ref{eq:Lambda_m}) to compute $\spai$. 
Shown in Table \ref{ktable} are the widely varying values of $k$  used in the different experiments.  Clearly the large  range of $k$-values, from 1.15 for EASTOP\cite{eastop} to 1.6 for Fly's Eye\cite{fly} differ significantly, thus making the {\em published} values of $\spai$ unreliable.  It is interesting to note the monotonic decrease over time in the $k$'s used in the different experiments, from 1.6 used in Fly's Eye in 1984 to the 1.15 value used in EASTOP in 2007, showing the time evolution of Monte Carlo models of energy dissipation in showers.  For comparison, Monte Carlo simulations  made by Pryke\cite{pryke} in 2001 of  several more modern shower models are also shown in Table \ref{ktable}. We see that even among modern shower models, the spread is still significant. The purpose of this letter is a proposal
 to minimize the impact of model dependence on the determination of $\spai$.
 
\begin{table}
\caption{A table of $k$-values, used in experiments and from Monte Carlo model simulation\label{ktable}} 
 \begin{tabular}[b]{|l||c|}
\hline
Experiment&k\\ \hline\hline
Fly's Eye&1.6\\ \hline
AGASSA&1.5\\ \hline
Yakutsk&1.4\\ \hline
EASTOP&1.15\\ \hline

     \multicolumn{2}{c}{Monte Carlo Results: C.L. Pryke}\\ 
      \hline
     Model&$k$ \\ \hline
     CORSIKA-SIBYLL&$1.15\pm 0.05$\\
          \hline
MOCCA--SIBYLL&$1.16\pm 0.03$\\
          \hline\hline
CORSIKA-QGSjet&$1.30\pm 0.04$\\
          \hline
MOCCA--Internal&$1.32\pm 0.03$\\
\hline\hline
\end{tabular}
\end{table}

{\em $\spai$ from the $X_{\rm max}$ distribution: Method II.} The HiRes group\cite{belov} has developed a quasi model-free method of measuring $\spai$ directly.  They fold into their shower development program a randomly generated exponential distribution of shower first interaction points, and then fit the entire distribution, and not just the trailing edge, as is done in the experiments of  Ref. \cite{fly,akeno,yakutsk,eastop}.  They obtain $\spai=460\pm14\ ({\rm stat})+39\ ({\rm syst})-11\ ({\rm syst})$ mb at $\sqrt s=77$ GeV, a result which they claim is effectively model-independent and hence is an absolute determination\cite{belov}.

\begin{figure}
\begin{center}
\mbox{\epsfig{file=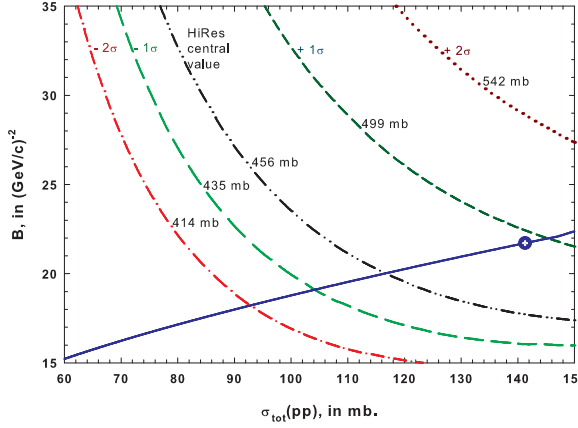%
              ,width=3in,bbllx=2pt,bblly=33pt,bburx=408pt,bbury=340pt,clip=%
}}
\end{center}
\caption[]{  $B$ dependence on the $pp$ total cross section $\sigma_{pp}$. The
 five curves are lines  of constant  $\spai$,  of 414, 435, 456, 499 and
 542 mb---the central value is the published Fly's Eye value, and the others
 are $\pm 1\sigma$ and $\pm 2\sigma$. The solid curve is a plot of a 
 QCD-inspired fit of $B$ against $\sigma_{pp}$, obtained from a $\ln^s$ fit---see text for details.  The large dot is the prediction for
$\spai$ at $\sqrt s=77$ TeV, the HiRes energy.}
\label{fig:p-air}
\end{figure}
{\em Extraction of $\sigma_{pp}$ from $\spai$}. The total $pp$ cross section is extracted from $\spai$ in two distinct steps. First, one calculates the $p$-air {\em total} cross section, $\spa$,  from
 the measured inelastic production cross section using 
\begin{equation}
\spai = \spa - \spae - \spaqe \,.  \label{eq:spa}
\end{equation}
Next, the Glauber method\cite{yodh} is used to transform the measured
 value of $\spai$ into a proton--proton total cross section $\sigma_{pp}$;
 all the necessary steps are calculable in the theory. In Eq.\,(\ref{eq:spa})
 the measured cross section for particle production is supplemented with
$\spae$ and $\spaqe$, 
 the elastic and quasi-elastic cross section, respectively, as calculated by the
 Glauber theory, to obtain the total cross section $\spa$. The subsequent
 relation between $\spai$ and $\sigma_{pp}$ critically involves the nuclear slope parameter $B$,  the logarthmic slope of forward elastic $pp$ scattering, ${d\sigma_{pp}^{\rm el}/ dt}$, i.e.,
\begin{equation}
B\equiv  \left[ {d\over dt} \left(\ln{d\sigma_{pp}^{\rm el}\over dt}\right)
 \right]_{t=0} \,,.
\end{equation}
 A plot of $B$ against
 $\sigma_{pp}$, 5 curves of different values of $\spai$, is shown in Fig.\,\ref{fig:p-air}, taking into account inelastic screening{\cite{engelBsig}. The reduction procedure
 from $\spai$ to $\sigma_{pp}$ is summarized in Ref. \cite{engel}.
 The solid curve in Fig.\,\ref{fig:p-air} is a plot of $B$  vs. 
 $\sigma_{pp}$, which we will discuss in detail later.

{\em Determination of $\sigma_{pp}(s)$}.  Block and Halzen\cite{bhfroissartnew} have made an analytic amplitude fit that saturates the Froissart bound\cite{froissart}, to both the available high energy total cross section and $\rho$-value data, where $\rho$ is defined as the ratio of the real to the imaginary portion of the forward scattering amplitude, for both $\bar pp$ and $pp$ interactions.  For their high energy expressions they used the analytic amplitude form
\ba
\sigma^{\pm}(\nu)&=&c_0+c_1\ln\left(\frac{\nu}{m}\right)+c_2\ln^2\left(\frac{\nu}{m}\right)+\beta_{\cal P'}\left(\frac{\nu}{m}\right)^{\mu -1}\nonumber\\
&&\pm\  \delta\left({\nu\over m}\right)^{\alpha -1},\label{sigmapmpp}\\
\rho^\pm(\nu)&=&{1\over\sigma^\pm(\nu)}\left\{\frac{\pi}{2}c_1+c_2\pi \ln\left(\frac{\nu}{m}\right)\right.\nonumber\\
&&\left.-\beta_{\cal P'}\cot({\pi\mu\over 2})\left(\frac{\nu}{m}\right)^{\mu -1}+\frac{4\pi}{\nu}f_+(0)\right.\nonumber\\
&&\left.\qquad\qquad\qquad\pm \delta\tan({\pi\alpha\over 2})\left({\nu\over m}\right)^{\alpha -1} \right\}\label{rhopmpp},
\ea
where the upper sign is for $pp$ and the lower sign is for  $\bar pp$ scattering, with $\mu=0.5$, $\nu$ is the laboratory energy, $f_+(0)$ is a dispersion relation subtraction constant, and  $m$ the proton mass. The 7 real constants $c_0,c_1,c_2,\beta_{\cal P'},\delta,\alpha$ and $f_+(0)$ are parameters of the fit. Since at high energies, $s$, the square of the cms  energy, is given by $2m\nu$, we see that their cross section approaches $\ln^2s$ at high energies, thus saturating the Froissart bound\cite{froissart}.    

Using all of the cross sections, $\sigma_{pp}$ and $\sigma_{\bar pp}$, along with all of the $\rho$-values,  $\rho_{\bar p p}$ and $\rho_{ p p}$,  in the Particle Data Group\cite{pdg} archive that were in the laboratory energy interval $18.3\le\nu\le 1.73\times 10^6$ GeV, i.e., $6\le\sqrt s\le 1800$ GeV, Block and Halzen\cite{bhfroissartnew} formed a sieved data set using the sieve algorithm of Ref. \cite{sieve} to eliminate outliers, which markedly improved their fit\cite{bhfroissartnew}.   Using 4 analyticity constraints\cite{blockanalyticity}, i.e., by fixing both the cross sections $\sigma_{\bar pp}$ and $\sigma_{pp}$ and their laboratory energy derivatives, at $\sqrt s=4$ GeV, they reduced the number of parameters to be fit from 7 to 4 and obtained an excellent fit, which, in turn, constrained $pp$ cross sections at cosmic ray energies to have a relative accuracy $\sim 1-2$\%. Their fits to $\sigma$ and $\rho$ are shown in Fig. \ref{fig:ppcurves}(a) and Fig. \ref{fig:ppcurves}(b), respectively.

\begin{figure}
\begin{center}
\mbox{\epsfig{file=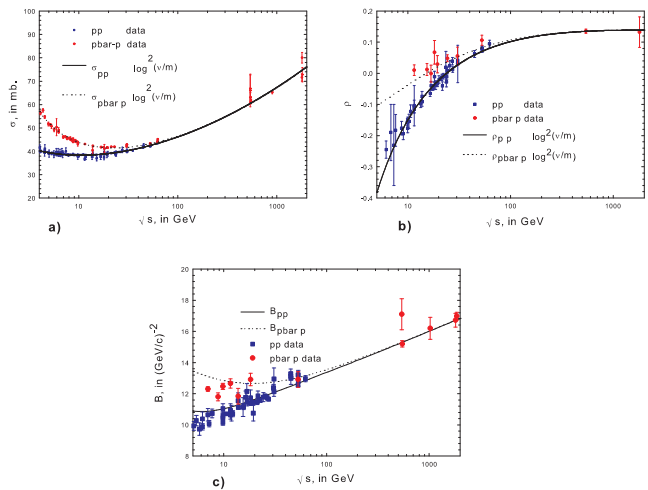%
            ,width=3.4in,bbllx=65pt,bblly=0pt,bburx=504pt,bbury=334pt,clip=%
}}
\end{center}
\caption[]{The saturated Froissart bound fit\cite{bhfroissartnew}  of total cross section
 $\sigma_{pp}$, $\rho$ {\em vs.} $\sqrt s$, in GeV, for $pp$
 (squares) and $\bar pp$ (circles) accelerator data:
 (a) $\sigma_{pp}$, in mb,\ \  (b) $\rho$;\ (c) the nuclear slope $B$,
 in GeV$^{-2}$ {\em vs.} $\sqrt s$, in GeV, from a QCD-inspired fit\cite{blockphysreports}.}
\label{fig:ppcurves}
\end{figure}

{\em Determination of $B$(s)}. A QCD-inspired parameterization\cite{blockphysreports}  of forward
 $\bar pp$ and $pp$ scattering amplitudes
 which is analytic, unitary and fits all data of $\sigma_{\rm tot}$, $B$
 and $\rho$ for both $\bar pp$ and $pp$ interactions has been made,  using 2 analyticity constraints which fix $\sigma_{\bar pp}$ and $\sigma_{pp}$ at $\sqrt s=4 $ GeV;  see Fig.\,\ref{fig:ppcurves}(c) for $B(s)$. 
 
The solid curve in Fig.\,\ref{fig:p-air} is a plot of $B$ vs. $\sigma_{pp}$, with $B$ taken from the QCD-inspired fit of Ref. \cite{blockphysreports} and  $\sigma_{pp}$ taken from the Froissart bound fit of Ref. \cite{bhfroissartnew}. The large dot corresponds to the value of $\sigma_{pp}$ and $B$ at $\sqrt s$ = 77 TeV, the HiRes energy, thus fixing the predicted value of $\spai$ at the HiRes energy.
 
 {\em Obtaining $\sigma_{pp}$ from  $\spai$}. In Fig.\,\ref{fig:sigpp_p-air}, we have plotted the values of
 $\sigma_{pp}$ vs. $\spai$ that are deduced from the
 intersections of the $B$-$\sigma_{pp}$ curve  with the $\spai$
 curves in Fig.\,\ref{fig:p-air}. Figure~\,\ref{fig:sigpp_p-air}
furnishes cosmic ray experimenters with an easy method to convert their measured $\spai$ to $\sigma_{pp}$, and vice versa. %
The percentage error in $\spai$ is $\approx 0.4$\% near $\spai = 450 $mb, due to the error 
in $\sigma_{pp}$ from  model parameter uncertainties.  

\begin{figure}[h]
\begin{center}
\mbox{\epsfig{file=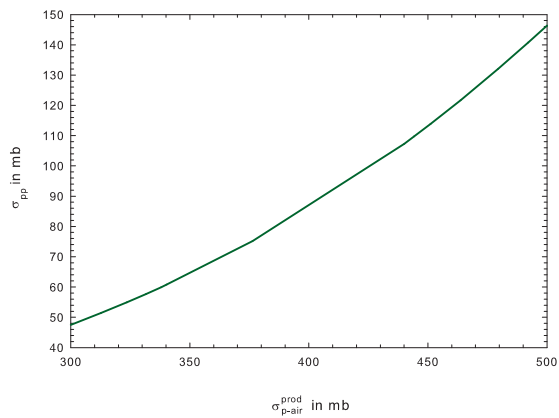%
            ,width=3in,bbllx=65pt,bblly=245pt,bburx=500pt,bbury=560pt,clip=%
}}
\end{center}
\caption[]{A plot of the predicted total pp cross section $\sigma_{pp}$, in mb
 {\em vs.} the measured p-air cross section, $\spai$, in mb.
}
\label{fig:sigpp_p-air}
\end{figure}

{\em Determining the $k$ value}.  It is important at this point  to recall Eq.\,(\ref{eq:Lambda_m}), $
\Lambda_m = k \lpa $, thus rewinding us of  the fact that in Method I, the extraction of $\lpa$ (or $\spai$) from the
 measurement of $\Lambda_m$ requires knowing the parameter
 $k$. The measured depth $X_{\rm max}$ at which a shower reaches
 maximum development in the atmosphere, which is the basis of the
 cross section measurement in Ref. \cite{fly}, is a combined measure
 of the depth of the first interaction, which is determined by
 the inelastic cross section, and of the subsequent shower development,
 which has to be corrected for. 
 The model dependent rate of shower development and its fluctuations
 are the origin of the deviation of $k$ from unity
 in Eq.\,(\ref{eq:Lambda_m}). As seen in Table \ref{ktable}, its values range from 1.6 for a very old model
 where the inclusive cross section exhibited Feynman scaling, to 1.15
 for modern models with large scaling violations.

 Adopting the same  strategy that earlier had been used by Block et al.\cite{blockhalzenstanev}, we decided to match the data to our prediction of $\spai (s)$ in order to extract
 a {\em common} value for $k$. This neglects the possibility
of a weak energy dependence of $k$ over the range measured, found to be very small in the simulations of Ref. \cite{pryke}.
By combining the results of Fig.\,\ref{fig:ppcurves}\,(a) and Fig.\,\ref{fig:sigpp_p-air}, we obtain our prediction of $\spai$  vs. $\sqrt s$, which is shown in Fig. \ref{fig:p-aircorrected2}.  To determine $k$, we leave it as a free parameter and make a $\chi^2$ fit to {\em  rescaled} $\spai (s)$ values of Fly's Eye, \cite{fly}AGASSA\cite{akeno}, EAS-TOP\cite{eastop} and Yakutsk\cite{yakutsk}, which are the experiments that need a common $k$-value. 
\begin{figure}[b]
\begin{center}
\mbox{\epsfig{file=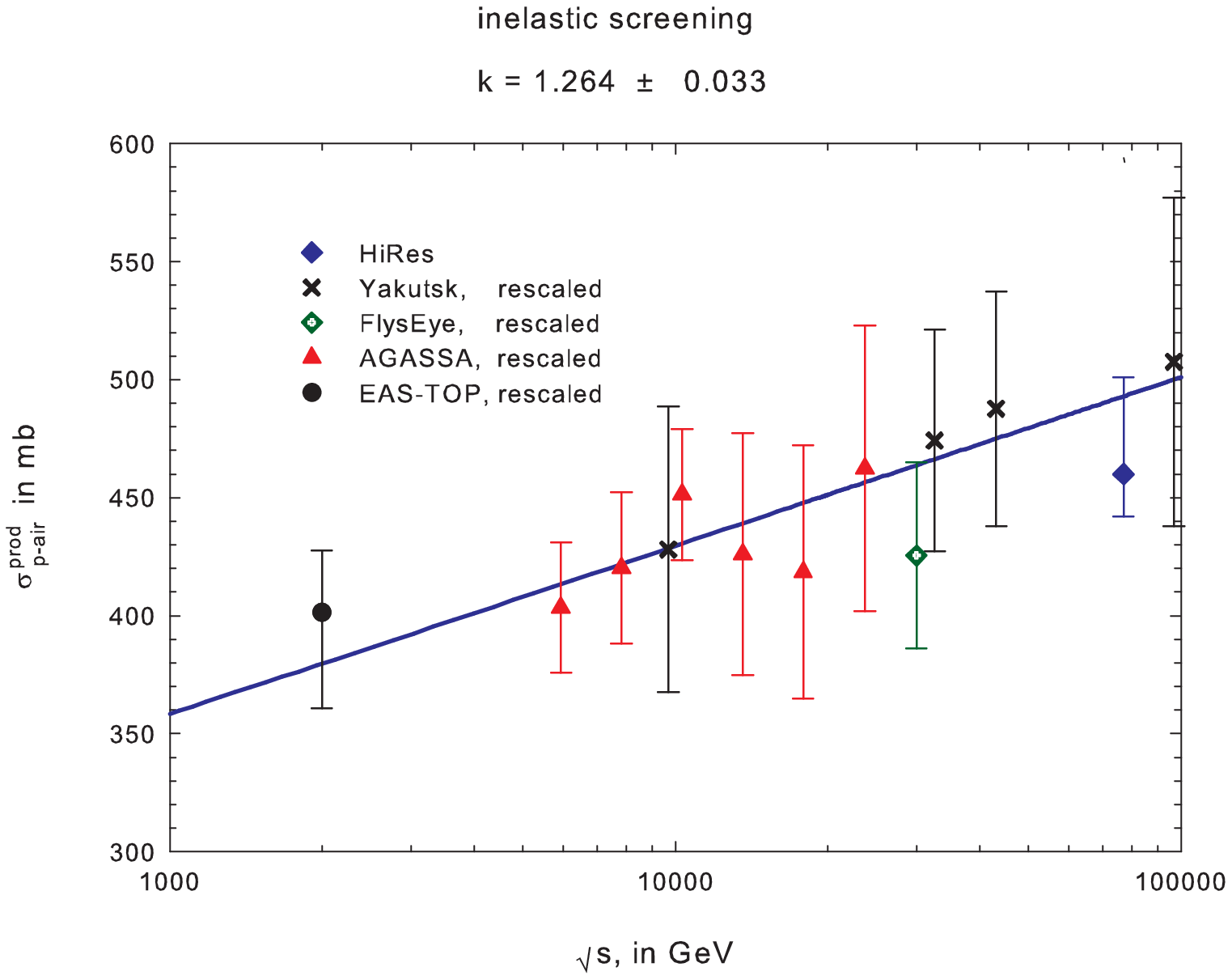%
              ,width=3in,bbllx=0pt,bblly=55pt,bburx=440pt,bbury=360pt,clip=%
}}
\end{center}
\caption[]{\protect
{  A $\chi^2$ fit of the {\em renormalized} AGASA, EASTOP, Fly's Eye and Yakutsk data for $\spai$, in mb,
 as a function of the energy, $\sqrt s$, in GeV. The result of the fit for the
 parameter $k$ in Eq. (\ref{eq:Lambda_m}) is $k=1.263\pm0.033$. The HiRes point (solid diamond), at $\sqrt s=77$ GeV, is the model-independent HiRes experiment, which has {\em not} been renormalized. 
}
}
\label{fig:p-aircorrected2}
\end{figure}

Figure \,\ref{fig:p-aircorrected2} is a plot of $\spai$ vs. $\sqrt s$, the cms energy in GeV, for the two different types of experimental extraction, using Methods I and II described earlier.  Plotted {\em as published} is the HiRes value at $\sqrt s=77$ TeV, since it is an absolute measurement. We have rescaled in Fig.\,\ref{fig:p-aircorrected2} the published values of $\spai$ for  Fly's Eye\cite{fly}, AGASSA\cite{akeno}, Yakutsk\cite{yakutsk} and EAS-TOP\cite{eastop},  against our prediction of $\spai$, using the {\em common} value of $k=1.264 \pm 0.033\pm 0.013$ obtained from a $\chi^2$ fit, and it is the rescaled values that are plotted in Fig. \ref{fig:p-aircorrected2}.
The error in $k$ of $0.033$ is the statistical error of the $\chi^2$ fit, whereas the error of $0.013$ is the systematic error due to the error in the prediction of $\spai$.     Clearly,
 we have an excellent fit, with complete agreement for  all experimental points. Our analysis gave  $\chi^2=3.19$ for 11 degrees of freedom (the low
 $\chi^2$ is likely due to overestimates of experimental errors).
 We note that our $k$-value,  $k=1.264\pm0.033\pm0.013$, is about halfway between the values of CORSIKA-SIBYLL and CORSIKA-QSGSjet found in the Pryke simulations\cite{pryke}, as seen in Table \ref{ktable}. 

We next compare our measured $k$ parameter with a direct measurement of $k$ by the HiRes group\cite{belovkfactor}. They measured the exponential slope of the tail of their $X_m$ distribution, $\Lambda_m$ and compared it to the p-air interaction length $\lpa$ that they found. Using \eq{eq:Lambda_m}, they deduced that $k=1.21+0.14-0.09$, in agreement with our value, giving us additional experimental confirmation of our method. 
 
{\em Conclusions}.  Our measured $k$ value, $k=1.264 \pm 0.033\pm 0.013$, agrees very well with the $k$-value measured by the HiRes group, at the several parts per mil level, and in turn, they both agree with Monte Carlo model simulations at the 5--10 part per mil level.  

It should be noted that the EASTOP\cite{eastop} cms energy, 2 TeV, is essentially  identical to the top energy of the Tevatron collider, where there is an {\em experimental} determination of $\sigma_{\bar pp}$\cite{comment}, and consequently, no necessity for an {\em extrapolation} of collider cross sections.  Since  their value of $\spai$  is in excellent agreement with the predicted value of $\spai$, this serves to anchor our fit at its low energy end. Correspondingly, at the high end of the cosmic ray spectrum, the absolute value of the HiRes experimental value of $\spai$ at 77 TeV---which requires {\em no knowledge} of the $k$ parameter---is also in good agreement with our prediction, anchoring the fit at the high end. Thus, our $\spai$ predictions, which span the enormous energy range, $2\le\sqrt s\le 100000$ TeV, are completely consistent with {\em all} of the cosmic ray data, for both magnitude and energy dependence.

In the future, we look  forward
 to the possibility of confirming our analysis with the high
 statistics of the Pierre Auger Collaboration\cite{Auger}, as well as confirming the prediction of $107.3\pm 1.2$ mb for the total $p p$ cross section\cite{bhfroissartnew} at the LHC
 energy of 14 TeV.

{\em Acknowledgments}. We would like to thank the Aspen Center for Physics for its
 hospitality during the  writing of this manuscript.

\end{document}